\documentclass[12pt]{jpconf}
\usepackage{amssymb}
\usepackage{amsmath}
\usepackage{mathrsfs}
\usepackage{bbm}
\usepackage[dvips]{graphicx}
\usepackage{psfrag}
\usepackage{multirow}
\usepackage{hyperref}

\usepackage{tikz}
\usetikzlibrary{matrix,positioning}

\numberwithin{equation}{section}

\newcommand{\be}{\begin{equation}}
\newcommand{\ee}{\end{equation}}
\newcommand{\ds}{\displaystyle}
\newcommand{\ts}{\textstyle}
\newcommand{\id}[1]{\mathbbm{1}_{#1}}
\newcommand{\iden}{\mathbbm{1}}

\makeatletter
\def\Ddots{\mathinner{\mkern1mu\raise\p@
\vbox{\kern7\p@\hbox{.}}\mkern2mu
\raise4\p@\hbox{.}\mkern2mu\raise7\p@\hbox{.}\mkern1mu}}
\makeatother

\begin{document}

\title
{A finite variant of the Toom Model}
\author{Arvind Ayyer}
\address{Department of Mathematics, \\
Indian Institute of Science,\\ 
Bangalore - 560012, India}
\ead{arvind@math.iisc.ernet.in}
\date{\today}

\begin{abstract}
We present results for a finite variant of the one-dimensional Toom model with closed boundaries. We show that the steady state distribution is not of product form, but is nonetheless simple. In particular, we give explicit formulas for the densities and some nearest neighbour correlation functions. We also give exact results for eigenvalues and multiplicities of the transition matrix using the theory of ${\mathscr R}$-trivial monoids in joint work with A. Schilling, B. Steinberg and N. M. Thi\'ery.
\end{abstract}

\section{Introduction}
Dynamical variants of the Ising model have been of interest since Glauber's pioneering work \cite{g1} on the subject. Although there are many dynamical rules for the Ising model whose steady state distribution is Gibbsian, there are many other dynamical rules whose steady state distribution has very different structure, see for example \cite{gls,glms}.

The model that we will be interested in is a probabilistic cellular automaton
of Ising spins on $\mathbb{Z}^2$ due to A. Toom \cite{toom}. The dynamics is 
given by the so-called NEC (north-east-center) rule. Recall that cellular automata are
discrete-time dynamical models, where the entire configuration is updated simultaneously. 
Denote the spin at
the site $(i,j)$ at integer time $t \in \mathbb{Z}_+$ as $\sigma_{i,j}(t)$, where  $\sigma_{i,j}(t)$ can be $\pm$. The dynamics then evolves as follows.
\[
\sigma_{i,j}(t+1) = \begin{cases}
+ & \text{with prob. p} \\
- & \text{with prob. q} \\
\text{sign}(\sigma_{i,j}(t) + \sigma_{i+1,j}(t) + \sigma_{i,j+1}(t)) & \text{with prob. 1-p-q}
\end{cases}
\]
To be more explicit, the spin at a given site at time $t+1$ evolves according to three spins at time $t$; itself, its north neighbour and its east neighbour. We illustrate the rule below, where we only show the case where $\sigma_{i,j}(t)=+$. The other case is similar. For clarity, we have circled the changing spin.
\[
\begin{array}{c c}
+ & \\
\oplus & +
\end{array}
\xrightarrow{q}
\begin{array}{c c}
+ & \\
\ominus & +
\end{array},
\quad
\begin{array}{c c}
+ & \\
\oplus & -
\end{array}
\xrightarrow{q}
\begin{array}{c c}
+ & \\
\ominus & -
\end{array},
\quad
\begin{array}{c c}
- & \\
\oplus & -
\end{array}
\xrightarrow{1-p}
\begin{array}{c c}
- & \\
\ominus & -
\end{array}
\]
Notice that the model becomes a deterministic NEC majority rule when $p=q=0$. If $p \neq q$, the rule favours one spin over the other, which can be thought of as the effect of a magnetic field.

It turns out that the model exhibits a noise-dependent phase transition, which was demonstrated rigorously by Toom \cite{toom}. He showed that for sufficiently high $p,q$, there is a unique steady state, whereas for low enough $p,q$, there are at least two steady states.
 
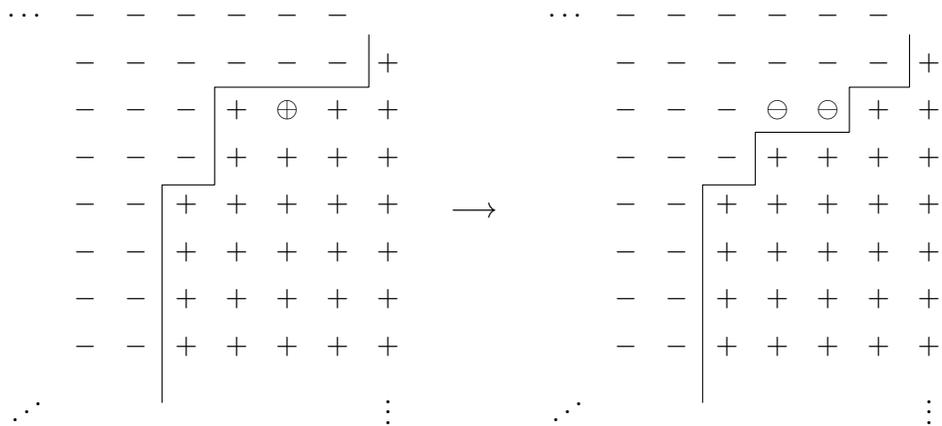
\begin{figure}[htbp!]
\begin{center}
\begin{tabular}{ccc}
\begin{tikzpicture}[scale=0.5,>=latex,line join=bevel]
\matrix (m) [matrix of math nodes,ampersand replacement=\&, row sep=.3em, column sep=.3em]
{\cdots \& - \& - \& - \& - \& - \& - \& \\
\& - \& - \& - \& - \& - \& - \& + \\
\& - \& - \& - \& + \& \oplus \& + \& + \\
\& - \& - \& - \& + \& + \& + \& + \\
\& - \& - \& + \& + \& + \& + \& + \\
\& - \& - \& + \& + \& + \& + \& + \\
\& - \& - \& + \& + \& + \& + \& + \\
\& - \& - \& + \& + \& + \& + \& + \\
\Ddots \& \&  \&  \&  \&  \&  \& \vdots \\
};
\draw  (4.4,4.8) -- (4.4,3.4) -- (0.3,3.4) -- (0.3,0.8) -- (-1.1,0.8) -- (-1.1,-5.0);
\end{tikzpicture} 
&
\raisebox{3cm}{$\longrightarrow$} 
&
\begin{tikzpicture}[scale=0.5,>=latex,line join=bevel]
\matrix (m) [matrix of math nodes,ampersand replacement=\&, row sep=.3em, column sep=.3em]
{\cdots \& - \& - \& - \& - \& - \& - \& \\
\& - \& - \& - \& - \& - \& - \& + \\
\& - \& - \& - \& \ominus \&  \ominus \& + \& + \\
\& - \& - \& - \& + \& + \& + \& + \\
\& - \& - \& + \& + \& + \& + \& + \\
\& - \& - \& + \& + \& + \& + \& + \\
\& - \& - \& + \& + \& + \& + \& + \\
\& - \& - \& + \& + \& + \& + \& + \\
\Ddots \& \&  \&  \&  \&  \&  \& \vdots \\
};
\draw (4.4,4.8) -- (4.4,3.4) -- (2.8,3.4) -- (2.8, 2.2) -- (0.3,2.2) -- (0.3,0.8) -- (-1.1,0.8) -- (-1.1,-5.0);
\end{tikzpicture}
\end{tabular}
\caption{Dynamics in the PCA at low temperature in the third quadrant. The $+$ spin which will spontaneously flip is shown on the left as $\oplus$ and the spins which flipped to $-$ are shown in the right as $\ominus$.}
\label{fig:isingdynamics}
\end{center}
\end{figure}

Derrida, Lebowitz, Speer and Spohn \cite{dlss1,dlss2} considered this model on the third quadrant in $\mathbb{Z}^2$, with the boundary condition that all spins on the negative $x$-axis are $-$ and all on the negative $y$-axis are $+$. When $p=q=0$, any configuration
in which $+$'s and $-$'s are separated by a single interface of staircase shape is stationary, see Figure~\ref{fig:isingdynamics}. They were interested in the dynamics of the interface for very small $p,q$. First, notice that in that case, all defects in the bulk of these phases are quickly cancelled out. However, the interface can and does fluctuate because of spontaneous sign changes of the $+$'s below and the $-$'s to the left of the interface. Figure~\ref{fig:isingdynamics} shows how this happens for a $+$ spin below the interface. Notice that the new interface is still of staircase shape.

Derrida, Lebowitz, Speer and Spohn \cite{dlss1,dlss2} understood the interface dynamics
by thinking of the interface as itself an Ising spin configuration $S$ on $\mathbb{Z}_+$ as follows. Look at the steps taken by the interface starting from the north-east corner.
If the $n$'th step is vertical, set $S_n = +$, otherwise set $S_n = -$.

The transition shown in Figure~\ref{fig:isingdynamics} corresponds to the transition
\be \label{ising1d-dlss}
+ - \underline{- - +} + - + + + + \cdots \quad \longrightarrow \quad + - \underline{+ - -} + - + + + + \cdots,
\ee
where the part undergoing the change has been underlined. In this one-dimensional model,
 each $\pm$ exchanges with the first $\mp$ on its right with rate $\lambda_{\pm}$.
They also proposed to call this one-dimensional version the ``Toom model''.

However, there turns out to be a simplification of this model due to Lebowitz, Neuhauser and Ravishankar (LNR) \cite{lnr}. Rather than allowing all spins to exchange, they only allow the leftmost spin in a block to exchange with the first opposite spin to its right. The transition in \eqref{ising1d-dlss} would not be allowed in this model. An example of an allowed transition is
\be \label{ising1d-lnr}
+ \underline{- - - +} + - + + + + \cdots \quad \longrightarrow \quad + \underline{+ - - -} + - + + + + \cdots.
\ee
Further, the first spin flips independently with a prescribed rate $\alpha$.  They studied the steady state distribution of this model when $\lambda_+ = \lambda_- = \alpha = 1$. The most interesting result there from our point of view is \cite[Theorem 2]{lnr} that the density at site $n$ goes like $\ds \frac{1}{2\sqrt{\pi n}}$.

It is this variant that we will refine.
We consider the model on an interval of $L$ sites with closed
boundaries. 
We modify the LNR model by introducing two parameters
$\alpha$ and $\beta$. In the original model \cite{lnr}, $\alpha$ and $\beta$ were
taken to be 1. We will also make the connection to exclusion processes explicit
by replacing the spins $\pm$ by $0,1$. Each site is 
then occupied either by a particle of type 0 or that of
type 1. The dynamics is as follows: the leftmost particle in a block
of particles of the same type can exchange with the opposite type
particle to the right of the block. Clearly, the site preceding the block must
be of opposite character if there is such a site.
This exchange occurs with rate
$\alpha$ if the block contains 0's and $\beta$ if the block contains
1's. Note that the block size can be one. 
To recap,
\be \label{bulkrules}
\begin{array}{ccl}
\underbrace{0 \dots 0}_k 1 \longrightarrow 1 \underbrace{0 \dots 0}_k, & \text{ with
  rate $\alpha$,} & \multirow{2}{*}{\text{if the block begins at the first site, and}}\\
\underbrace{1 \dots 1}_k 0 \longrightarrow 0 \underbrace{1 \dots 1}_k, & \text{ with
  rate $\beta$,} & \\
1\underbrace{0 \dots 0}_k 1 \longrightarrow 11 \underbrace{0 \dots 0}_k, & \text{ with
  rate $\alpha$,} & \multirow{2}{*}{\text{if the block does not begin at the
  first site.}}\\
0\underbrace{1 \dots 1}_k 0 \longrightarrow 00 \underbrace{1 \dots 1}_k, & \text{ with
  rate $\beta$,} &
\end{array}
\ee
An example of this dynamics is given with different notation in \eqref{ising1d-lnr}.
The model is particle conserving and one defines it
using integers $(n_0,n_1)$, with $n_0+n_1=L$, to denote the number of
holes and particles respectively. 
There are thus $\binom{n_0+n_1}{n_0}$ configurations. 
We will call this the $(n_0,n_1)$-system.
The rules then are just given by
\eqref{bulkrules} for any positive value of $k$. In particular, the
rightmost block cannot by itself make any transitions. It is of course
affected by the transitions involving the block immediately to its left.

Note that the presence of the boundary is crucial to having a
nonequilibrium steady state. If we consider the model on a ring of $L$
sites, then there is a unique absorbing set of configurations with all $n_0$
holes in one block and all $n_1$ particles in another block. These
configurations are translations of one another, and the steady state
is uniformly distributed on this small set of configurations. 
On the interval model, however, the boundary acts to break up the size
of blocks. For example, in the $(3,3)$ system, if one starts with the
configuration $001110$, then one can have a transition to the
configuration $100110$ with rate $\alpha$. Then one has split the 1's
from a block of size 3 to two blocks.

We will present results for this model in the rest of the paper. The proofs will be delegated to a subsequent publication \cite{ayyer1}. In Section~\ref{sec:matrix}, we demonstrate the structure of the Markov matrices for the model. We will first give a recurrence for constructing the matrices, and give formulas for their eigenvalues.
In Section~\ref{sec:corr}, we present formulas for the density and other nearest neighbour correlations. We will also present a conjecture for the partition function.

\section{The Markov Matrices} 
\label{sec:matrix}
Here we will show that the Markov matrices for the model have a rich structure. Recall that the Markov matrix of the $(n_0,n_1)$-system is a matrix of size $\binom{n_0+n_1}{n_0}$ indexed by configurations of the system whose $(i,j)$'th entry is the rate of going from state $j$ to state $i$. We will denote the matrix by $M_{n_0,n_1}$.

As an illustration, consider the example with $n_0 = n_1 = 2$. The configurations, in lexicographic order are: $(0011, 0101, 0110, 1001, 1010, 1100)$. 
The Markov matrix in this ordered basis is given by
\be \label{mat22}
M_{2,2} = 
\left( 
\begin {array}{cccccc} 
-\alpha&\beta&\beta&0&0&0\\
\noalign{\medskip}0&-\beta-2\alpha&0&\beta&0&0\\
\noalign{\medskip}0&\alpha&-\beta-\alpha&0&\beta&\beta\\
\noalign{\medskip}\alpha&\alpha&0&-\beta-\alpha&\beta&0\\
\noalign{\medskip}0&0&\alpha&0&-2\beta-\alpha&0\\
\noalign{\medskip}0&0&0&\alpha&\alpha&-\beta
\end {array} 
\right)
\ee
The steady state distribution is the column (or right) eigenvector with eigenvalue 0. We will normalise it so that the sum of the entries is 1, i.e. it represents a probability distribution. We will denote the vector by $\pi_{n_0,n_1}$. For this case,
\be \label{statdist22}
\pi_{2,2} = 
\left( 
{\frac {{\beta}^{2}}{ \left( \alpha+\beta \right) ^{2}}},
{\frac {\alpha{\beta}^{3}}{ \left( \alpha+\beta \right) ^{4}}},
{\frac {{\alpha}^{2}\beta\, \left( \alpha+2\,\beta \right) }{ \left( \alpha+\beta \right) ^{4}}},
{\frac {\alpha{\beta}^{2}\, \left( 2\,\alpha+\beta \right) }{ \left( \alpha+\beta \right) ^{4}}},
{\frac {{\alpha}^{3}\beta}{ \left( \alpha+\beta \right) ^{4}}},
{\frac {{\alpha}^{2}}{ \left( \beta+\alpha \right) ^{2}}}
\right)
\ee
Note the simple probabilities of configurations with just two clusters.
We will also see later that $M_{n_0,n_1}$ has simple formulas for eigenvalues and their multiplicities. The eigenvalues of $M_{2,2}$ are
\be \label{eig22}
\begin{matrix}
0 & \text{with multiplicity } 1, \\
-\alpha - \beta & \text{with multiplicity } 4, \\
-2\alpha - 2\beta & \text{with multiplicity } 1.
\end{matrix}
\ee

We will begin by explaining the recursive structure of the Markov matrices.
To that end, we will first need to define some
auxiliary matrices. Let $\id{k}$ denote the $k \times k$ identity
matrix and let $A_{n_0,n_1}$ be a matrix of size
$\binom{n_0+n_1-2}{n_1-1} \times \binom{n_0+n_1-1}{n_1}$ defined as a
concatenation of identity matrices of increasing sizes which are
justified at the top row as,
\be \label{defA}
A_{n_0,n_1} = \left( \begin{array}{c c c c}
\id{\binom{n_1-1}{n_1-1}} & \id{\binom{n_1}{n_1-1}} & \dots &
\id{\binom{n_0+n_1-2}{n_1-1}}
\end{array} \right),
\ee
and the rest of whose entries are zero. For example,
\[
A_{3,3} = \left( \begin{array}{c c c c c c c c c c}
1 & 1 & 0 & 0 & 1 & 0 & 0 & 0 & 0 & 0 \\
0 & 0 & 1 & 0 & 0 & 1 & 0 & 0 & 0 & 0 \\
0 & 0 & 0 & 1 & 0 & 0 & 1 & 0 & 0 & 0 \\
0 & 0 & 0 & 0 & 0 & 0 & 0 & 1 & 0 & 0 \\
0 & 0 & 0 & 0 & 0 & 0 & 0 & 0 & 1 & 0 \\
0 & 0 & 0 & 0 & 0 & 0 & 0 & 0 & 0 & 1
\end{array} \right).
\]
Similarly, let $B_{n_0,n_1}$ be a matrix of size
$\binom{n_0+n_1-2}{n_0-1} \times \binom{n_0+n_1-1}{n_0}$ defined as a
concatenation of identity matrices of increasing sizes which are
justified at the bottom row,
\be \label{defB}
B_{n_0,n_1} = \left( \begin{array}{c c c c}
\id{\binom{n_0+n_1-2}{n_0-1}} & \dots & \id{\binom{n_0}{n_0-1}} &
\id{\binom{n_0-1}{n_0-1}}
\end{array} \right),
\ee
and the rest of whose entries are zero. For example,
\[
B_{3,3} = \left( \begin{array}{c c c c c c c c c c}
1 & 0 & 0 & 0 & 0 & 0 & 0 & 0 & 0 & 0 \\
0 & 1 & 0 & 0 & 0 & 0 & 0 & 0 & 0 & 0 \\
0 & 0 & 1 & 0 & 0 & 0 & 0 & 0 & 0 & 0 \\
0 & 0 & 0 & 1 & 0 & 0 & 1 & 0 & 0 & 0 \\
0 & 0 & 0 & 0 & 1 & 0 & 0 & 1 & 0 & 0 \\
0 & 0 & 0 & 0 & 0 & 1 & 0 & 0 & 1 & 1
\end{array} \right).
\]

We will write down the Markov matrix in the lexicographically ordered basis of configurations, just as in the example \eqref{mat22}.
The transition matrix $M_{n_0,n_1}$ can be expressed in terms of
$M_{n_0-1,n_1}$ and $M_{n_0,n_1-1}$ using $2 \times 2$ blocks as,
\[
M_{n_0,n_1} = \left( \begin{array}{c |c}
M^{(0,0)} & M^{(0,1)} \\[0.1cm]
\hline \\[-0.3cm]
M^{(1,0)} & M^{(1,1)}
\end{array} \right),
\]
where each block can be expressed in terms of smaller blocks as
\[
\begin{split}
M^{(0,0)} &= M_{n_0-1,n_1} - \alpha \left( \begin{array}{c}
0 \\
\hline
A_{n_0,n_1}
\end{array} \right), \\
M^{(0,1)} &= \beta \left( \begin{array}{c}
0 \\
\hline
B_{n_0,n_1}
\end{array} \right), \\
M^{(1,0)} &= \alpha \left( \begin{array}{c}
A_{n_0,n_1} \\
\hline
0
\end{array} \right), \\
M^{(1,1)} &= M_{n_0,n_1-1} - \beta \left( \begin{array}{c}
B_{n_0,n_1} \\
\hline
0
\end{array} \right).
\end{split}
\]
Here $0$ represents a zero matrix, $A_{n_0,n_1}$ and $B_{n_0,n_1}$ are defined in \eqref{defA} and \eqref{defB} respectively and with the initial condition $M_{0,n_1} = M_{n_0,0} = (0)$, the $1 \times 1$ zero matrix.
The sizes of the blocks of zero matrices is fixed by the fact that
$M^{(0,0)}$ and $M^{(1,1)}$ are square matrices of size
$\binom{n_0+n_1}{n_1}$ and  $\binom{n_0+n_1}{n_0}$ respectively.
One can check that the matrix of the example at the beginning of this section in \eqref{mat22} can be recovered from the recursion above. The proof of this recursion follows from an elementary analysis of transitions in the ordered basis. For instance, $M^{(0,1)}$ represents all transitions from configurations that begin with a 1 to those that begin with a 0.

The deepest results that have been obtained about this model are on the eigenvalues and their multiplicities of the Markov matrices in joint work with A. Schilling, B. Steinberg and N. M. Thi\'ery \cite{asst-new}. The result there is more general, but it reduces to
the following in our case.
The characteristic polynomial of the transition matrix $M_{n_0,n_1}$
is given by the explicit product formula
\be \label{charpoly}
| \lambda \iden - M_{n_0,n_1}| = \prod_{k=0}^{\ts \min(n_0,n_1)}
(\lambda +k(\alpha+\beta))^{\ts \binom{n_0}{k} \binom{n_1}{k}},
\ee
where the factor for $k=0$ corresponds to the zero eigenvalue, which naturally occurs with multiplicity one. One can check that the eigenvalues for $n_0 = n_1 = 2$ in \eqref{charpoly} match those given in \eqref{eig22}.

The proof of \eqref{charpoly} is highly nontrivial and 
has been obtained using techniques from the representation
theory of ${\mathscr R}$-trivial monoids; 
another application of this theory is to a nonabelian model of sandpiles \cite{asst2}. 
No elementary proof of \eqref{charpoly} has been found so far.

\section{Correlation Functions}
\label{sec:corr}
We will now state exact results for some correlation functions including the density. In some cases, the analysis is straightforward and will be explained, but the more complicated cases will be stated without proof. Detailed proofs will appear elsewhere \cite{ayyer1}.

We will use $\eta_i$ to denote the occupation variable of the particle at site $i$ and angle brackets $\langle \cdot \rangle$ to denote averages in the steady state distribution. As before, we will set $n_0 + n_1 = L$.
First of all, note that the system satisfies an obvious particle-hole symmetry. Namely
$\langle \eta_1, \dots,\eta_L \rangle$ in the $(n_0,n_1)$-system is equal to
$\langle 1-\eta_1, \dots,1-\eta_L \rangle$ in the $(n_1,n_0)$-system with $\alpha$ and $\beta$ interchanged.

It turns out that the nature of explicit formulas for correlation functions is 
different for the sites less than both $n_0$ and $n_1$.
Consequently, it will be useful to denote $n^* = \min(n_0,n_1)$.
This is because looking at the conditional process on the first $n^*$ sites is
equivalent to looking at the first $n^*$ sites in the semi-infinite
system \cite{lnr} because no matter how large a block of particles (holes) is
within this subsystem, there is always a hole (particle) far enough to
the right.

First of all, the density at the first site is given by
\be \label{dens1}
\langle \eta_1 \rangle = \frac{\alpha}{\alpha+\beta}.
\ee
This is clear since there is a hole somewhere in the system for an
exchange with the leftmost site (which is of course the leftmost site
for the first block),
\[
0 = \frac d{dt} \langle \eta_1  \rangle = \alpha \langle 1-\eta_1
\rangle  - \beta \langle \eta_1 \rangle.
\]
With just a little more work, we can also show that the density in the first $n^*$ sites is constant, i.e.,
\be \label{dens<n*}
\langle \eta_k \rangle = \frac{\alpha}{\alpha+\beta},
\ee
for $1 \leq k \leq n^*$.
To see this, suppose that $2 \leq k \leq n^*$, and consider the master equation for $\langle \eta_k \rangle$. First, we look at the outgoing contribution. If  $\eta$  is a configuration with  $\eta_k = 1$  then, if  $\eta_{k-1} = 1$  then  $\eta_k$  cannot change instantaneously. If  $\eta_{k-1} = 0$, then $\eta_k$ can become $0$ in two ways: at rate  $\alpha$ due to an exchange with $\eta_{k-j}$  for some $j \geq  1$ and at rate  $\beta$  due to an exchange with $\eta_{k+j}$  for some $j \geq  1$. 
Thus the total negative contribution to  $\frac d{dt} \langle \eta_k
\rangle$ is
\[
- ( \alpha + \beta ) \langle (1 - \eta_{k-1}) \eta_k \rangle.
\]
Similarly, if  $\eta$  is a configuration with  $\eta_k = 0$  then, if  $\eta_{k-1} = 0$  then  $\eta_k$  cannot change instantaneously. If  $\eta_{k-1} = 1$ then $\eta_k$ can become $1$ in two ways: at rate  $\beta$  due to an exchange with $\eta_{k-j}$  for some $j \geq  1$ and at rate  $\alpha$  due to an exchange with $\eta_{k+j}$  for some $j \geq 1$.
Thus the total positive contribution to  $\frac d{dt} \langle \eta_k \rangle$  is
\[
( \alpha + \beta ) \langle \eta_{k-1} (1-\eta_k) \rangle.
\]
So
\be
\begin{split}
  \frac d{dt} \langle \eta_k \rangle
   &=   (\alpha+\beta)[-\langle(1-\eta_{k-1})\eta_k\rangle+
   \langle\eta_{k-1}(1-\eta_k)\rangle] \\ 
   &= (\alpha+\beta)[\langle\eta_{k-1}\rangle - \langle\eta_k\rangle],
\end{split}
\ee
and this must vanish in the steady state. Using the initial condition
in \eqref{dens1} proves \eqref{dens<n*}.

The reason the above proof does not go through if $k>n^*$ is that
it assumes there is a particle of the opposite type far enough to the
right of the block starting at $k$. When $k$ is too large, such an
assumption does not hold.
The density at a general site is given by the more complicated formula,
\be \label{densgen}
\langle \eta_{k} \rangle = 
\frac {{\beta}^{k-n_1}}{ \left( \alpha+\beta \right)^{k}}
 \left( \sum _{j=0}^{L-k}
   \binom{k-1}{n_1-j-1}{\alpha}^{n_1-j}{\beta}^{j}
   +  \sum_{j=0}^{n_1-k-1}\binom{k}{j} {\beta}^{n_1-j}
   {\alpha}^{j} \right).    
\ee
A little bit of manipulation shows that \eqref{densgen} reduces to \eqref{dens<n*} when $k \leq n^*$. The proof of this result is more complicated and uses results for other correlations stated in \eqref{lastk} and \eqref{lastk-1}. 

For more complicated correlation functions such as block size distributions, we have
some preliminary results. For $k \leq n_1$, 
\be \label{firstk}
\langle \eta_1 \dots \eta_k \rangle =
\left( \frac{\alpha}{\alpha+\beta} \right)^k.
\ee
To see this, first consider the outgoing transitions for a configuration $\eta$
with $\eta_1 = \cdots = \eta_k = 1$. Note that $\eta_2,\dots,\eta_k$ cannot change instantaneously. The only position which can effect a transition is the first one, 
where a hole to the right of $k$ exchanges with the particle at the first site with rate $\beta$. The only incoming transitions to $\eta$ are from configurations $\eta'$ in which  $\eta'_1  = \cdots =  \eta'_{k-1} =1$ and $\eta'_k = 0$ is nonzero. There is a transition to $\eta$ at rate $\alpha$ which exchanges a particle to the right of $k$ with the hole at the $k$th site. 
Therefore
\[
0 = \frac d{dt} \langle \eta_1 \dots \eta_k \rangle = \alpha \langle
\eta_1 \dots \eta_{k-1} (1-\eta_k) \rangle  - \beta \langle \eta_1 \dots \eta_k
\rangle,
\]
which implies the recursion,
\[
\langle \eta_1 \dots \eta_k \rangle = \frac{ \alpha \langle \eta_1
  \dots \eta_{k-1}  \rangle }{\alpha+\beta}.
\]
Using \eqref{dens1} as the initial condition gives the desired result.

Our final results are about correlations at a block of sites at the end of the system.
The probability of having a block of $L+1-k$ 1's starting at position $k$ is given by
\be \label{lastk}
\langle \eta_{k} \cdots \eta_L \rangle =
\left(\frac{\beta}{\alpha+\beta} \right)^{k-1} \;\;
\sum_{i=0}^{k-1-n_0}  \binom{k-1}i
\left( \frac{\alpha}{\beta} \right)^i.
\ee
The formula becomes much simpler if we also look at the joint correlation with a hole at site $k-1$.
\be \label{lastk-1}
\langle (1-\eta_{k-1}) \eta_{k} \cdots \eta_L \rangle =
\binom {k-2}{n_0-1} \frac{\alpha^{k-1-n_0} \beta^{n_0}}{(\alpha+\beta)^{k-1}}.
\ee
Note that both formulas \eqref{lastk} and \eqref{lastk-1} give 0 when we set $k= n_0$
as expected. These will be proved in a subsequent publication \cite{ayyer1}. One can get the joint correlation of blocks of holes in the first and the last sites by interchanging $\alpha$ and $\beta$ in \eqref{firstk}, \eqref{lastk} and \eqref{lastk-1}. We do not yet have exact formulas for block correlations in the bulk.

Steady state probabilities are rational functions of $\alpha$ and $\beta$, and hence there is no unique way to write them. Suppose we write the steady state probabilities such that the greatest common factor of the numerators is 1. This forces them to be written in a unique way. See for example, \eqref{statdist22}.
We will call the {\em normalization factor}, denoted $Z(n_0,n_1)$, in the $(n_0,n_1)$-system to be the least common multiple of the denominators of the steady state probabilities when they are written in this way.
The formula for the normalization factor $Z(n_0,n_1)$ seems to be tantalisingly clean,
\be \label{partfn}
Z(n_0,n_1) = (\alpha + \beta)^{n_0 n_1}.
\ee
When $n_0 = n_1 = 2$, this matches with \eqref{statdist22}.
Although this formula seems quite simple, the proof has resisted the best of our efforts, and we leave it as a conjecture for now.
 
To conclude, let us compare some of our results with those of Lebowitz, Neuhauser and Ravishankar \cite{lnr}. First of all, our model is on the finite one-dimensional lattice, whereas theirs is on $\mathbb{Z}_{\geq 0}$. We have a more general model in the sense that we have more parameters. Note that the ratio of $\alpha$ and $\beta$ is the only extra parameter here, but in our generalised model \cite{asst-new}, we have many more.
We have explicit formulas for many correlation functions in \eqref{densgen},\eqref{firstk},\eqref{lastk-1},\eqref{lastk}, and a conjecture for the partition function \eqref{partfn}.

One can compare the densities in both models by performing asymptotics of \eqref{densgen}.
Of course, when $\alpha = \beta$ and $k$ is large, the density at site $k$ goes exactly like $\ds \frac{1}{2\sqrt{\pi k}}$, exactly as in \cite{lnr}.
A preliminary analysis seems to indicate that the density decays like $1/k$ for all values of $\alpha/ \beta$. 

\section*{Acknowledgements}
This work is partially supported by UGC Centre for Advanced Studies.
We thank J. L. Lebowitz and E. R. Speer for discussions during the initial stages of the project and an anonymous referee for helpful comments.

\section*{References}
\bibliographystyle{BibTeX/iopart-num/iopart-num.bst}
\bibliography{domwals}

\end{document}